\documentclass[prl,twocolumn,amsmath,amssymb,superscriptaddress,showpacs]{revtex4}

\usepackage{bm}
\usepackage{graphicx}
\newcommand{\ket}[1]{\lvert #1 \rangle}

\newcommand{\bracket}[1]{\langle #1 \rangle}
\newcommand{\BB}{\mathcal B}
\newcommand{\BA}{\mathcal A}
\newcommand{\BF}{\mathcal F}
\DeclareMathOperator{\IM}{Im}

\begin{document}

\title{Half-Heusler Compounds as a New Class of Three-Dimensional
  Topological Insulators}

\author{Di Xiao}

\affiliation{Materials Science \& Technology Division, Oak Ridge
  National Laboratory, Oak Ridge, TN 37831, USA}

\author{Yugui Yao}

\affiliation{Beijing National Laboratory for Condensed Matter Physics
  and Institute of Physics, Chinese Academy of Sciences, Beijing
  100190, China}

\affiliation{Department of Physics, The University of Texas at Austin,
  Austin, TX 78712, USA}

\author{Wanxiang Feng}

\affiliation{Beijing National Laboratory for Condensed Matter Physics
  and Institute of Physics, Chinese Academy of Sciences, Beijing
  100190, China}

\author{Jun Wen}

\affiliation{Department of Physics, The University of Texas at Austin,
  Austin, TX 78712, USA}

\author{Wenguang Zhu}

\affiliation{Department of Physics and Astronomy, The University of
  Tennessee, Knoxville, TN 37996, USA}

\affiliation{Materials Science \& Technology Division, Oak Ridge
  National Laboratory, Oak Ridge, TN 37831, USA}

\author{Xingqiu Chen}

\affiliation{Materials Science \& Technology Division, Oak Ridge
  National Laboratory, Oak Ridge, TN 37831, USA}

\author{G. Malcolm Stocks}

\affiliation{Materials Science \& Technology Division, Oak Ridge
  National Laboratory, Oak Ridge, TN 37831, USA}

\author{Zhenyu Zhang}

\affiliation{Materials Science \& Technology Division, Oak Ridge
  National Laboratory, Oak Ridge, TN 37831, USA}

\affiliation{Department of Physics and Astronomy, The University of
  Tennessee, Knoxville, TN 37996, USA}

\affiliation{ICQD, University of Science and Technology of China,
  Hefei, Anhui, 230026, China}

\begin{abstract}
Using first-principles calculations within density functional theory,
we explore the feasibility of converting ternary half-Heusler
compounds into a new class of three-dimensional topological insulators
(3DTI). We demonstrate that the electronic structure of unstrained
LaPtBi as a prototype system exhibits distinct band-inversion feature.
The 3DTI phase is realized by applying a uniaxial strain along the
[001] direction, which opens a bandgap while preserving the inverted
band order.  A definitive proof of the strained LaPtBi as a 3DTI is
provided by directly calculating the topological $\mathbb Z_2$
invariants in systems without inversion symmetry.  We discuss the
implications of the present study to other half-Heusler compounds as
3DTI, which, together with the magnetic and superconducting properties
of these materials, may provide a rich platform for novel quantum
phenomena.
\end{abstract}


\maketitle

Recent years have seen a surge of interest in a new class of materials
called topological insulators~\cite{qi2010,kane2010,moore2010}.  These
materials are distinguished from ordinary insulators by nontrivial
topological invariants associated with the bulk electronic
structure~\cite{moore2007,fu2007,roy2009}.  The existence of the
topological invariants dictates that the excitation gap must vanish at
the boundaries of a topological insulator, resulting in the formation
of robust metallic surface states.  A number of spectacular quantum
phenomena have been predicted when the surface states are under the
influence of magnetism and
superconductivity~\cite{qi2008,fu2008,fu2009,tanaka2009,yokoyama2010}.
To fully explore these phenomena thus demands great versatility from
the host material.  However, so far the experimental realizations of
topological insulators are limited to a few classes of simple
materials, including HgTe quantum well~\cite{bernevig2006,konig2007},
Bi$_{1-x}$Sb$_x$ alloy~\cite{fu2007a,hsieh2008}, and tetradymite
semiconductors such as Bi$_2$Se$_3$, Bi$_2$Te$_3$, and
Sb$_2$Te$_3$~\cite{zhang2009,xia2009,chen2009}.  Realizing the
necessary conditions for the predicted phenomena in these materials
can be difficult.

The search for topological insulators has greatly benefited from the
topological band theory.  It has been shown that for all known
topological insulators, in addition to the strong spin-orbit coupling,
their electronic structure can be characterized by a band-inversion
which involves the switching of bands with opposite parity around the
Fermi level~\cite{bernevig2006,fu2007a,zhang2009}.  This is very
similar to the quantum Hall effect in which two bands are allowed to
exchange their Chern numbers only when they come into contact with
each other~\cite{avron1983}.  The above observation suggests small
bandgap semiconductors or semimetals with heavy elements as promising
candidates as these materials are likely to develop an inverted band
order.

In this Letter we predict a new class of topological insulators
realized in small bandgap ternary half-Heusler
compounds~\cite{xiao2009a}.  In particular, using first-principles
calculations we demonstrate that LaPtBi as a prototype system becomes
a strong topological insulator upon the application of a uniaxial
strain along the [001] direction.  This result is first discussed
using the aforementioned band-inversion mechanism, then verified by
direct calculation of the $\mathbb Z_2$ topological invariants from
the bulk band structure~\cite{fu2006,fukui2007}.  Remarkably, ternary
half-Heusler compounds already boast an impressive list of the much
desired properties such as magnetism~\cite{canfield1991} and
superconductivity~\cite{goll2008}.  Together with the predicted
topological properties, these materials provide an exciting platform
for novel quantum phenomena.

\begin{figure}[b]
\includegraphics[width=6cm]{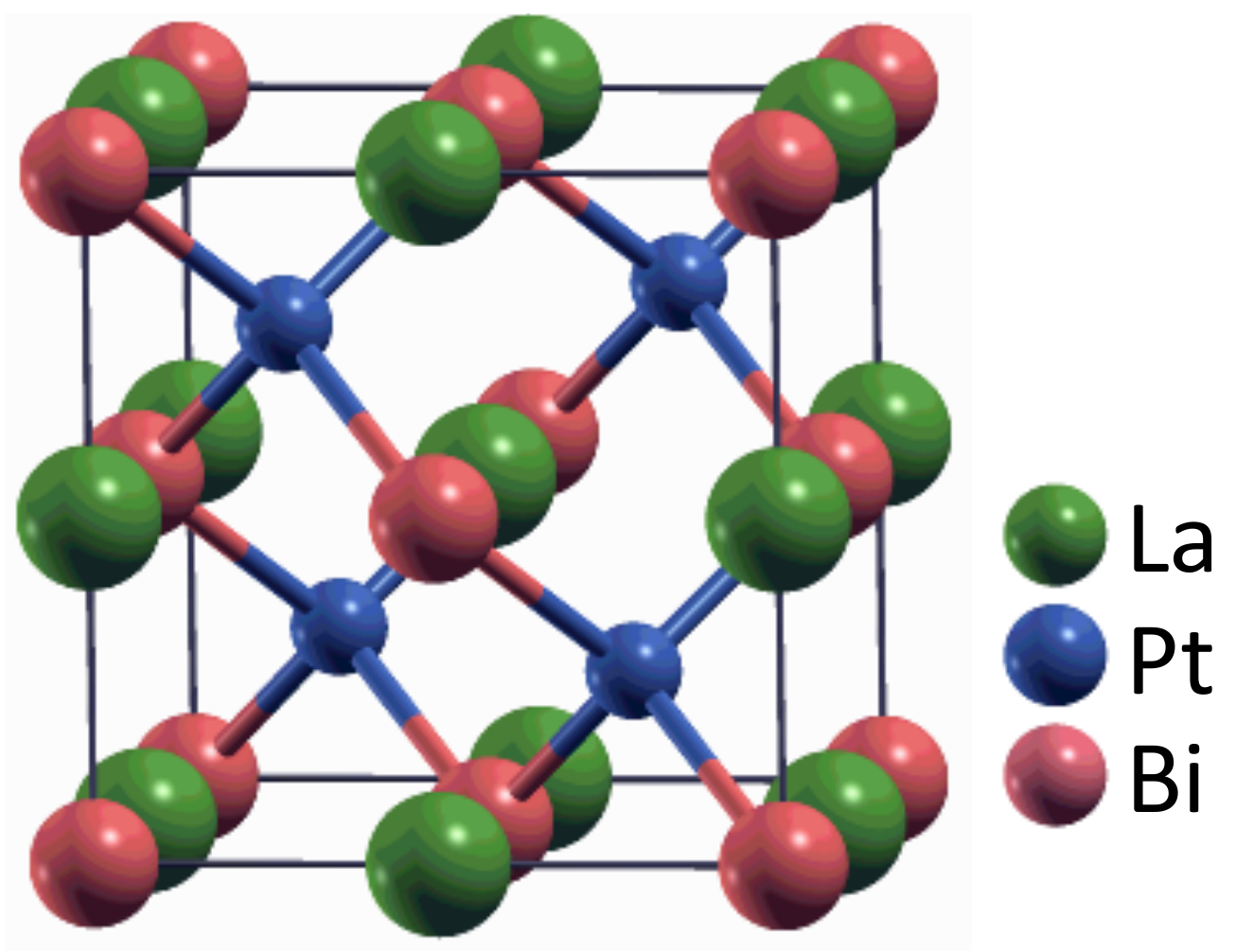}
\caption{\label{fig:struct}(color online) Crystal structure of
  half-Heusler compound LaPtBi in the $F\bar{4}3m$ space group. Green
  spheres at (0.5,0.5,0.5) are atom La, dark blue spheres at
  (0.25,0.25,0.25) are atom Pt, and pink spheres at (0,0,0) are atom
  Bi.}
\end{figure}

Ternary half-Heusler compounds have the chemical formula $XYZ$, where
$X$ and $Y$ are transition or rare earth metals and $Z$ a heavy
element.  Figure~\ref{fig:struct} shows the crystal structure of
LaPtBi, which consists of three interpenetrating, face-centered-cubic
lattices with Pt sitting at the unique site.  It can be regarded as a
hybrid compound of LaBi with the rock-salt structure, and LaPt and
PtBi with the zinc-blende structure.  Unlike tetradymite
semiconductors, the spatial inversion symmetry is broken in
half-Heusler structure.

\begin{figure}
\includegraphics[width=8cm]{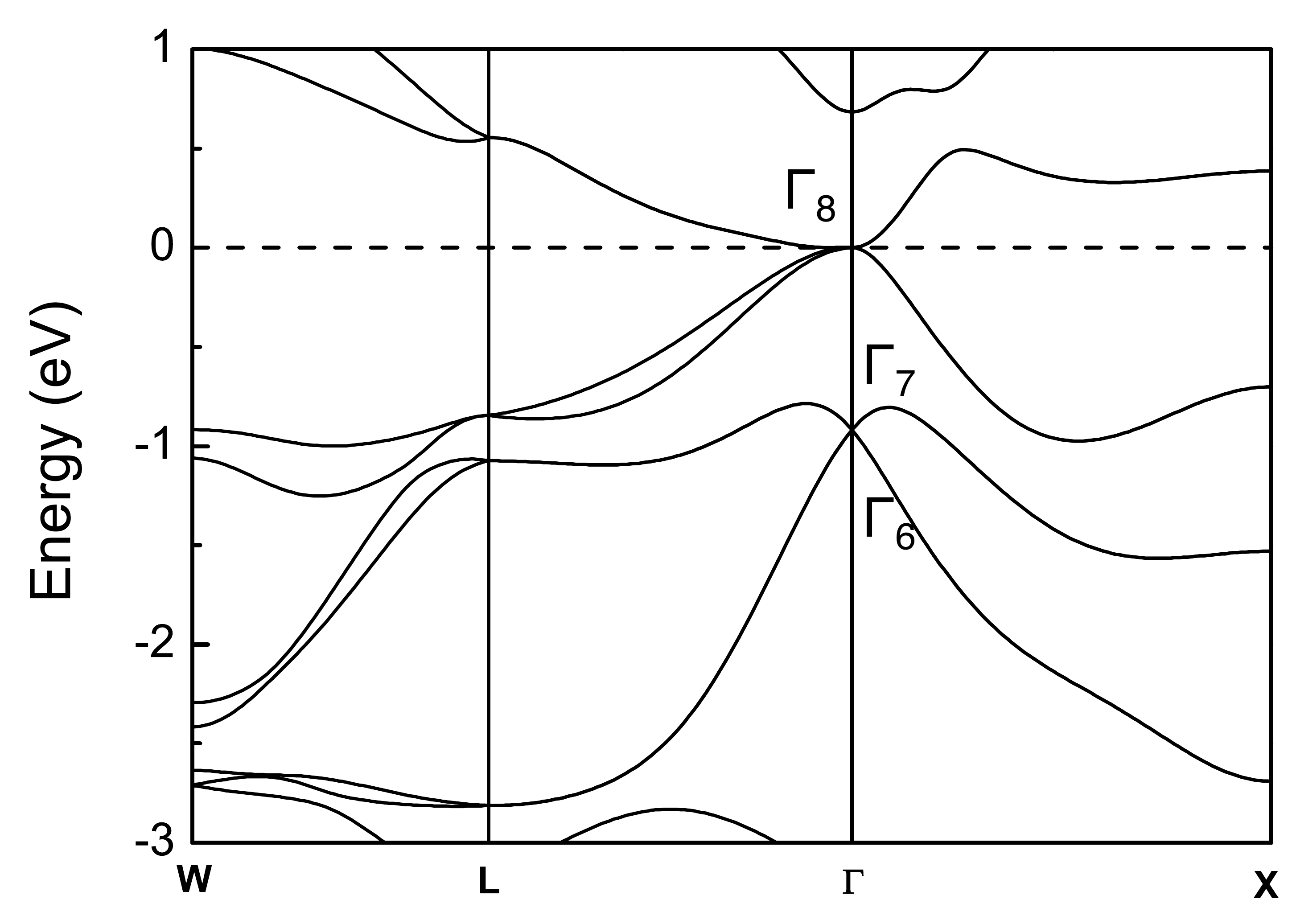}
\caption{\label{fig:band}Band structure of LaPtBi with experimental
  lattice constant a=6.83 \AA. The band order at the $\Gamma$ point is
  $\Gamma_8$, $\Gamma_7$, and $\Gamma_6$ in descendent order of
  energy. There is a small gap between the $\Gamma_7$ and $\Gamma_6$
  states.}
\end{figure}

When the total valence electron count in a primitive unit cell is 18,
the half-Heusler compounds are expected to have a band gap.  However,
some compounds display a distinctive semimetal behavior with LaPtBi
being one of the examples~\cite{jung2001}.  To investigate the band
topology, we employ the full-potential linearized augmented plane-wave
method~\cite{singh1994} with the local spin density approximation for
the exchange-correlation potential~\cite{perdew1992}.  Fully
relativistic band calculations were performed using the program
package \textsc{wien2k}~\cite{wien2k}. A converged ground state was
obtained using 10,000 $k$ points in the first Brillouin zone and
$K_\text{max}R_{MT} = 9$, where $R_{MT}$ represents the muffin-tin
radius and $K_\text{max}$ the maximum size of the reciprocal-lattice
vectors.  Wave functions and potentials inside the atomic sphere are
expanded in spherical harmonics up to $l = 10$ and 4, respectively.
Spin orbit coupling are included by a second-variational
procedure~\cite{singh1994}, where states up to 9 Ry above Fermi energy
are included in the basis expansion, and the relativistic $p_{1/2}$
corrections were also considered for 5p, 6p of Pt, and 6p of Bi in
order to improve the accuracy~\cite{kunes2001,larson2003}.  The
calculations were performed using the experimental lattice constant of
6.83 \AA~\cite{haase2002}.  Figure~\ref{fig:band} shows the fully
relativistic energy band structure of LaPtBi.  As anticipated from
experiments~\cite{jung2001}, LaPtBi is a semimetal with very small
electron and hole pockets around the $\Gamma$ point.  Our result is
consistent with previous calculations~\cite{oguchi2001}.

\begin{figure}
\includegraphics[width=8cm]{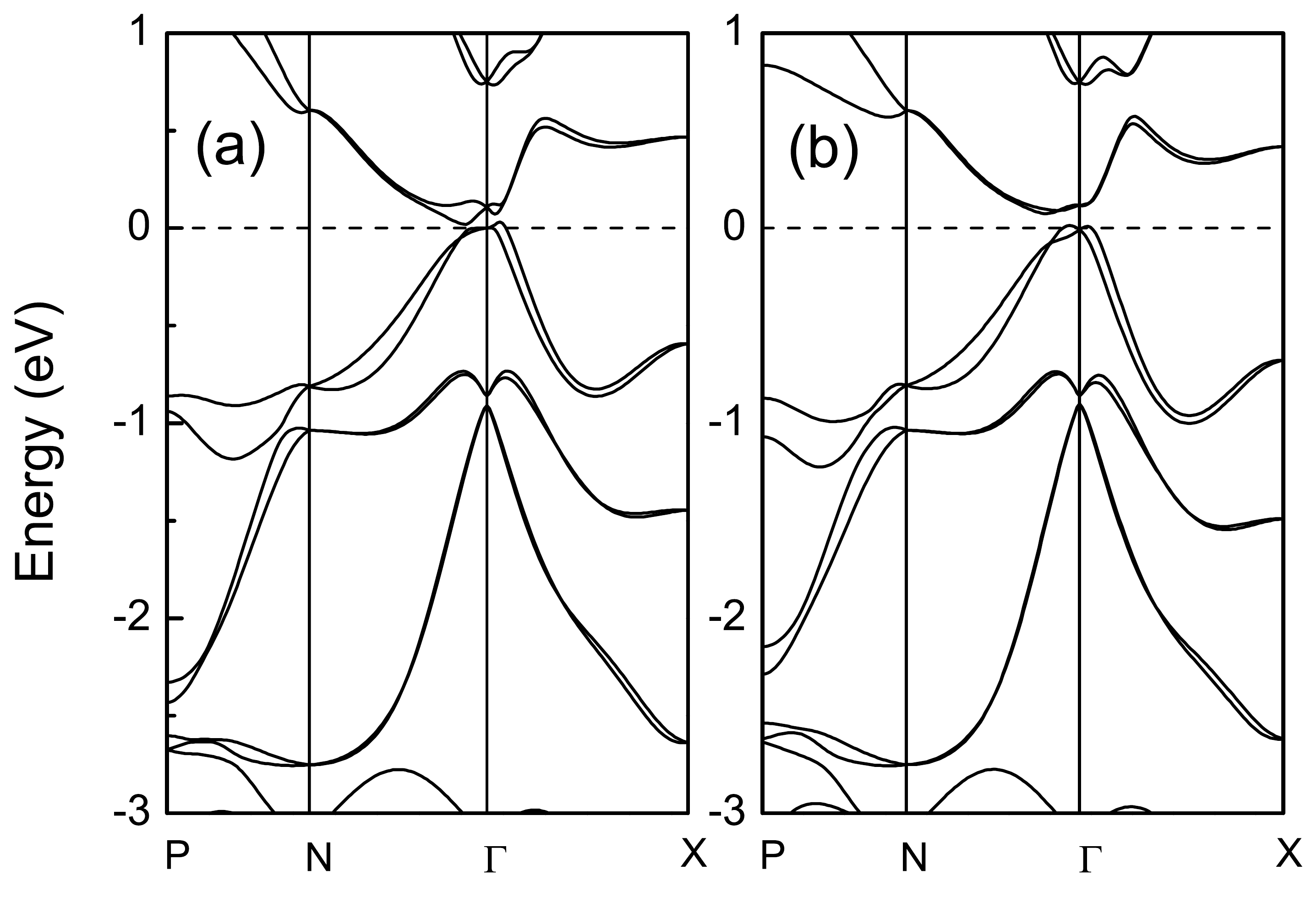}
\caption{\label{fig:uniaxial}Band structure of LaPtBi under uniaxial
  strain with constant volume along [001]-direction, reduce the c/a
  ratio by 5\% in (a), and increase the c/a ratio by 5\% in (b).  }
\end{figure}

\begin{figure}
\includegraphics[width=8cm]{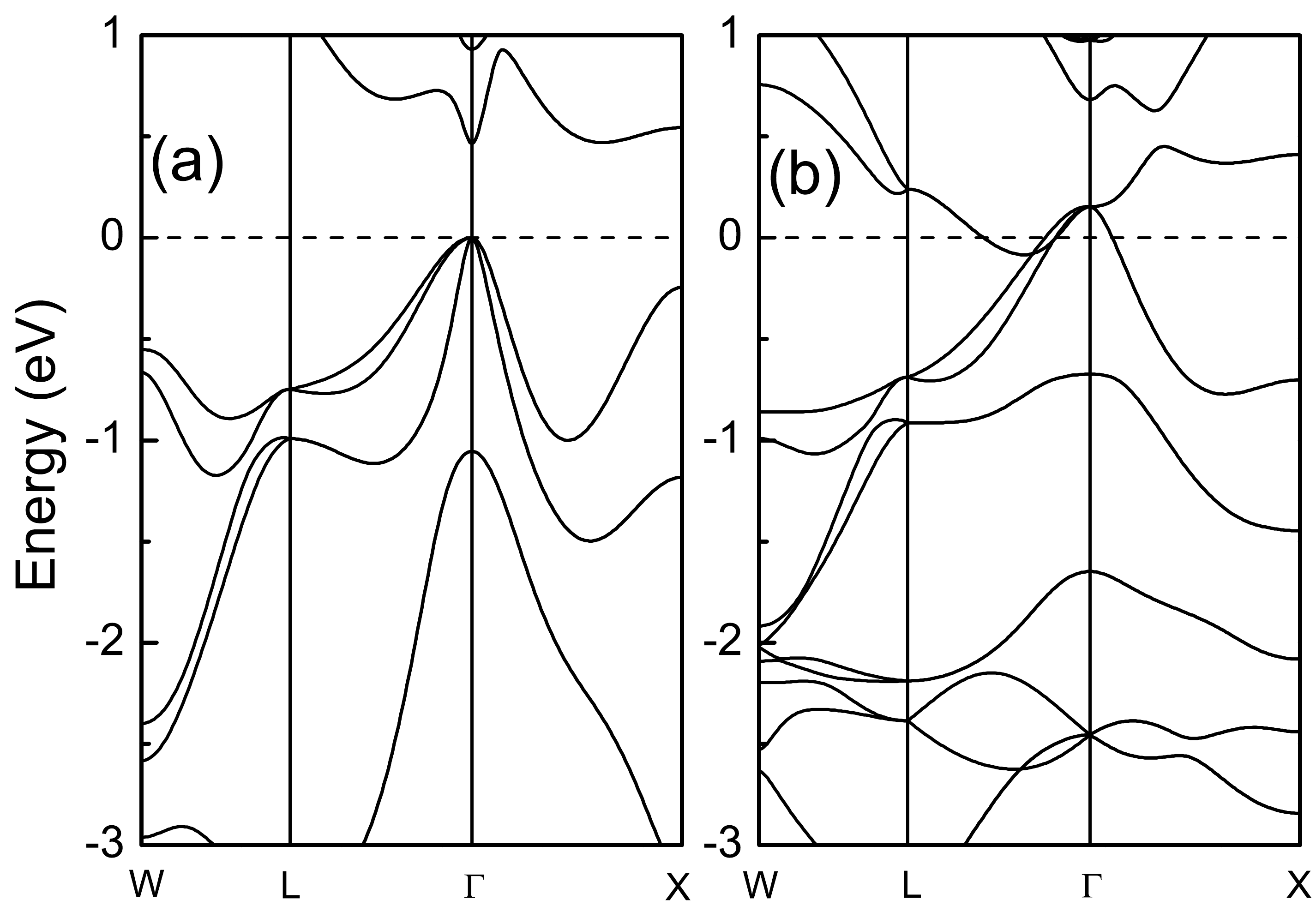}
\caption{\label{fig:uniform}Band structure of LaPtBi under hydrostatic
  strain, a=a-7\%a in (a), a=a+7\%a in (b)}
\end{figure}

As already pointed out by Og{\"u}chi~\cite{oguchi2001}, the band
structure near the Fermi level at the $\Gamma$ point is determined
mainly by PtBi with the zinc-blende structure, and the La states
participate in the band structure additively.  This separation allows
us to draw direct comparison with a known topologically nontrivial
compound HgTe~\cite{bernevig2006,fu2007a}, which is a II-VI material
also with the zinc-blende structure.  Let us focus on the bands at the
$\Gamma$ point close to the Fermi level.  Similar to the case of HgTe,
symmetry analysis shows that the fourfold degenerate $\Gamma_8$ states
lies above the twofold degenerate $\Gamma_7$ and $\Gamma_6$ states.
As discussed by Fu and Kane~\cite{fu2007a}, such a band inversion is a
strong indication that LaPtBi is in a topologically nontrivial state.
To remove the semi-metallic behavior, we apply a uniaxial strain along
the $[001]$ direction with constant volume to break the fourfold
degeneracy of the $\Gamma_8$ states.  Figure~\ref{fig:uniaxial} shows
the resulting band structure.  We find that although a local bandgap
can be opened, the system responds differently to compression and
elongation: Upon compression the material remains a semimetal while it
becomes an insulator when stretched.  The inverted band order stays
the same.  For comparison, we also calculated the band structure with
hydrostatic strain, shown in Fig.~\ref{fig:uniform}.  A global band
gap is opened by the hydrostatic strain when the lattice is compressed
while the semi-metallic behavior is retained with expansion.  However,
in the former case the $\Gamma_6$ states now jumps above the
$\Gamma_8$ states with the Fermi energy lies in between.  In this
situation, the material should be in a topologically trivial phase.

Although the band inversion near the $\Gamma$ point is a strong
indication that LaPtBi under uniaxial strain is in a topologically
insulating phase, it is not definitive because the topological
invariant is a global property of the entire Brillouin zone.  Fu and
Kane have proposed a parity criterion to identify topological insulators in systems
with both time-reversal and spatial inversion symmetry~\cite{fu2007a}.
However, it cannot be applied here because of the lack of inversion
symmetry in the half-Heusler structure.  Instead, we calculate the
topological invariants directly from the bulk band structure.

We first briefly describe the formalism for a 2D system.  In the
presence of time-reversal symmetry, Kramer's theorem dictates that the
energy eigenstates must come in pairs.  This allows us to enforce the
so-called time-reversal constraint on the Bloch functions:
\begin{equation} \label{constraint}
\ket{u_n(-\bm k)} = \Theta \ket{u_n(\bm k)} \;,
\end{equation}
where $\ket{u_n(\bm k)}$ is the periodic part of the Bloch function,
and $\Theta = e^{i\pi S_y/\hbar}K$ is the time reversal operator with
$S_y$ the spin operator and $K$ the complex conjugation.  Accordingly,
we only need to obtain Bloch functions in half of the Brillouin zone,
denoted by $\BB^+$, as those in the other half are fixed by
Eq.~\eqref{constraint}.  The band topology is characterized by the
$\mathbb Z_2$ invariant, given by~\cite{fu2006}
\begin{equation} \label{D}
\mathbb Z_2 = \frac{1}{2\pi} \Bigl[
\oint_{\partial\BB^+} d\bm k \cdot \bm{\BA}(\bm k) 
- \int_{\BB^+} d^2k\, \BF(\bm k) \Bigr]\text{ mod 2} \;,
\end{equation}
where $\bm{\BA}(\bm k) = i\sum_n\bracket{u_n(\bm k)|\bm\nabla_{\bm
    k}u_n(\bm k)}$ is the Berry connection and $\BF(\bm k) =
\bm\nabla_{\bm k} \times \bm{\BA}(\bm k)|_z$ is the Berry curvature;
the sum is over occupied bands. A topological insulator is
characterized by $\mathbb Z_2 = 1$ while ordinary insulators have
$\mathbb Z_2 = 0$.  The nonzero $\mathbb Z_2$ invariant is an obstruction to
smoothly defining the Bloch functions in $\BB^+$ under the
time-reversal constraint.

To numerically perform the integration, we follow the recipe by Fukui
and Hatsugai~\cite{fukui2007}.  The Bloch functions $\ket{u_n(\bm k)}$
are first obtained on a $\bm k$-space mesh in $\cal B^+$.  The mesh
must include the four time-reversal invariant $\bm k$-points: $\bm 0$,
$\bm G_1/2$, $\bm G_2/2$ and $(\bm G_1 + \bm G_2)/2$, expressed in
terms of the reciprocal lattice vectors.  After applying the
time-reversal constraint, next we introduce the link variable central
to many Berry-phase related
calculations~\cite{king-smith1993,resta1994}, given by $U_{\bm\mu}(\bm
k_j) = \det||\bracket{u_n(\bm k_j)|u_m(\bm k_j+\bm \mu)}||$, where
$\bm\mu$ is the unit vector on the mesh, and $n$ and $m$ run through
occupied bands.  The finite element expressions for $\BA$ and $\BF$
are $\BA_{\bm\mu}(\bm k_j) = \IM\log U_{\bm\mu}(\bm k_j)$, and
$\BF(\bm k_j) = \IM \log U_{\bm\mu}(\bm k_j) U_{\bm\nu}(\bm
k_j+\bm\mu) U^{-1}_{\bm\mu}(\bm k_j+\bm\nu) U^{-1}_{\bm\nu}(\bm k_j)$,
where the return value of the complex logarithm function is confined
to its principal branch $(-\pi, \pi]$.  We can then insert these
  expressions into Eq.~\eqref{D} to calculate the $\mathbb Z_2$
  invariant.  Define an integer field $n(\bm k_j)$ for each plaquette:
\begin{equation}
n(\bm k_j) = \frac{1}{2\pi}\Bigl\{[
\Delta_{\bm\nu}\BA_{\bm\mu}(\bm k_j) 
- \Delta_{\bm\mu}\BA_{\bm\nu}(\bm k_j)] -\BF(\bm k_j)\Bigr\}\;,
\end{equation} 
where $\Delta_{\mu}$ is the forward difference operator. The $\mathbb
Z_2$ invariant is given by the sum of the $n$-field in \emph{half} of
the Brillouin zone~\cite{fukui2007}, i.e., $\mathbb Z_2 = \sum_{\bm
  k_j \in \BB^+} n(\bm k_j) \mod 2$.

\begin{figure}
\includegraphics[width=\columnwidth]{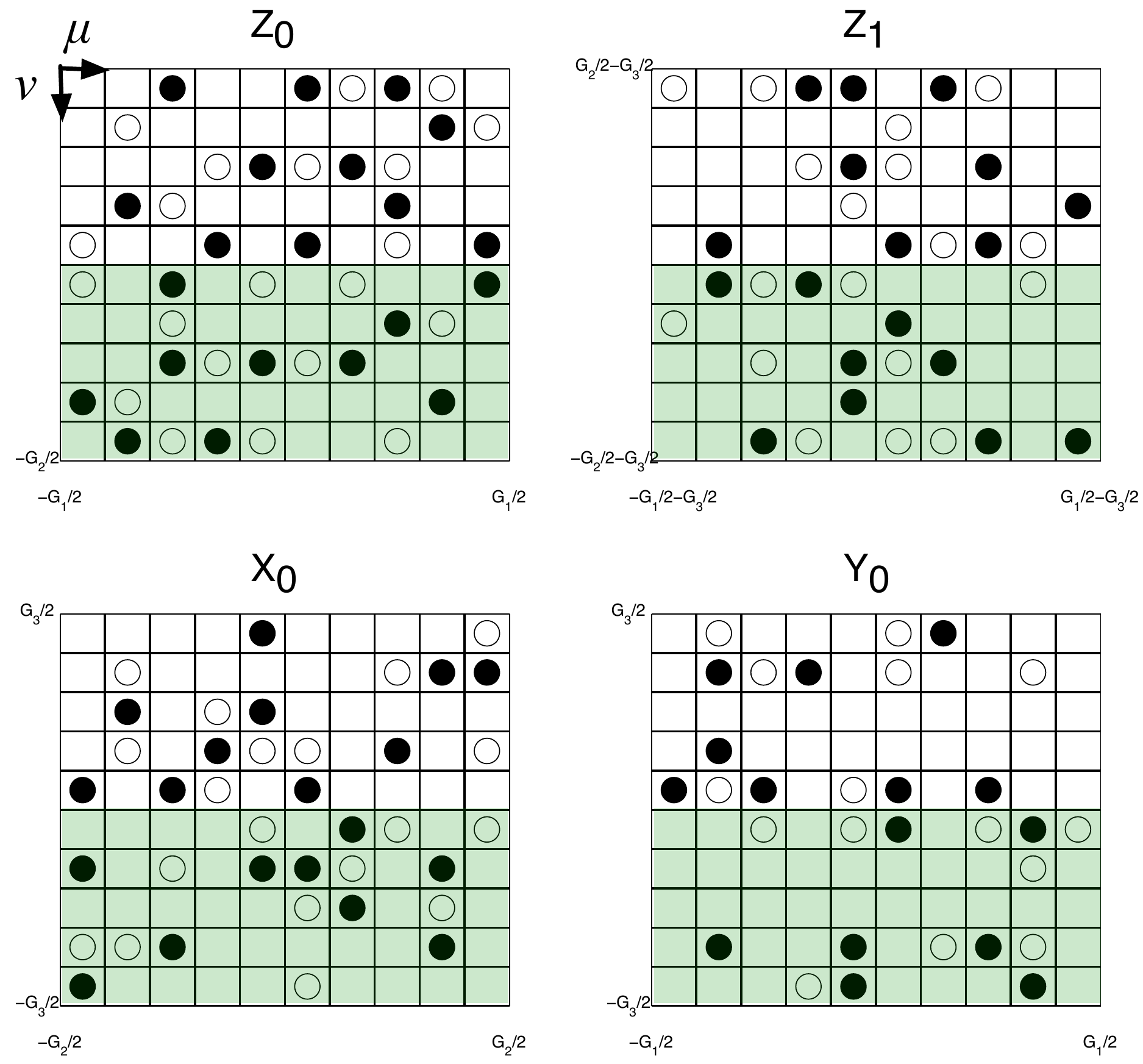}
\caption{\label{fig:Z21}The $n$-field configuration for LaPtBi under
  uniaxial strain computed under the time-reversal constraint.  The
  four tori are $Z_0$, $Z_1$, $X_0$ and $Y_0$ with the shaded area
  indicating half of the area.  $\bm\mu$ and $\bm \nu$ are the unit
  vectors of the $\bm k$-space mesh.  The white and black circles
  denote $n = 1$ and $-1$, respectively, while the blank denotes
  0. The $\mathbb Z_2$ invariant for each individual tori is obtained
  by summing the $n$-field over half of the tori.  These read $z_0=1$,
  $z_1=0$, $x_0=1$, and $y_0=1$. The $\mathbb Z_2$ invariants of the
  system are 1;(000).}
\end{figure}

In 3D the topology of the bands are characterized by
four independent $\mathbb Z_2$ invariants~\cite{fu2007,moore2007}.
These can be computed by considering six tori in the Brillouin zone.
For example, the torus $Z_0$ is spanned by $\bm G_1$ and $\bm G_2$
with the third component fixed at $0$, and $Z_1$ is obtained by fixing
the third component at $\bm G_3/2$.  The other four tori $X_0$, $X_1$,
$Y_0$, and $Y_1$ are defined similarly.  For each torus, one can calculate 
the corresponding $\mathbb Z_2$ invariant using the steps outlined above for 
2D systems.  Out of the six possible $\mathbb Z_2$ invariants only four of 
them are independent.  Following Ref.~\cite{fu2007,moore2007}, we use the 
notation $\nu_0;(\nu_1\nu_2\nu_3)$, with $\nu_0 = (z_0+z_1)\mod 2$, $\nu_1 = 
x_1$, $\nu_2 = y_1$ and $\nu_3 = z_1$, where $z_0$ is the $\mathbb Z_2$ 
invariant associated with the 2D torus $Z_0$.  The other $\mathbb Z_2$ invariants are defined similarly.  A nonzero $\nu_0$ indicates that the system is a strong topological insulator.

Figure~\ref{fig:Z21} shows the $n$-field configuration for LaPtBi
under uniaxial strain from first-principles calculations.  The
corresponding band structure is shown in Fig.~\ref{fig:uniaxial}(b).
Note that although the $n$-field itself depends on the gauge choice,
their sum over half of the Brillouin zone is gauge-invariant module
2.~\footnote{The sum of $n$-field over the entire Brillouin zone gives
  the Chern number, which vanishes identically for time-reversal
  invariant systems.} We find that LaPtBi under uniaxial strain
becomes a strong topological insulator with the topological invariants
1;(000).  However, under hydrostatic strain
[Fig.~\ref{fig:uniform}(a)], LaPtBi becomes an ordinary insulator.
Furthermore, although the band structure in Fig.~\ref{fig:uniaxial}(a)
shows a semi-metallic phase, we can still calculate the $\mathbb Z_2$
invariants for the bands because a local energy gap separates the conduction
and valance bands throughout the Brillouin zone.  In this case, LaPtBi becomes a topological metal
with $\nu_0 = 1$.

Having firmly established that LaPtBi under uniaxial strain realizes a
topological insulating phase, we have calculated a number of other
half-Heusler ternary compounds by first-principles method.  We find
that LuPtSb, ScPtBi, YPdBi, YPtSb have inverted band structure, and
the 3DTI phase can be induced by uniaxial strain along
[001]-direction. Generally, for many small bandgap half-Heusler
compounds, the topologically insulating phase can be realized by a
combination of hydrostatic strain to change the band order and
uniaxial strain to open an energy gap.  Details of the
first-principles calculations will be reported
elsewhere~\cite{feng2010}.

In conclusion, we have shown that the 18-electron ternary half-Heusler
compounds can be tuned into a new class of three-dimensional
topological insulators via proper strain engineering.  This is
confirmed by first-principles calculation of the topological $\mathbb Z_2$
invariants in systems without inversion symmetry.  This quantum
nature, plus other interesting physical properties of these materials,
such as magnetism~\cite{canfield1991} and
superconductivity~\cite{goll2008}, characterize these materials as an
exciting platform for novel quantum phenomena.

Note added: After the completion of the bulk of this work (see, e.g.,
the brief announcement in Ref.~\cite{xiao2009a}), two more related
studies have appeared~\cite{lin2010,chadov2010}, confirming and
expanding the predictions of the present work.

DX acknowledges useful discussions with Ying Ran.  
This work was supported by the Division of Materials Sciences and
Engineering, Office of Basic Energy Sciences, U.S. Department of
Energy, by NSF of China (10674163, 10974231), the MOST Project of
China (2006CB921300, 2007CB925000), and by Welch Foundation (F-1255).

\end{document}